# The pH Dependence of Ultrafast Charge Dynamics in Graphene Oxide Dispersions

**Georgia Kime[1,2], Kai-Ge Zhou[3,4], Samantha J. O. Hardman[5], Rahul R. Nair[3,4], Konstantin Sergeevich Novoselov[2], Daria V. Andreeva[6], and David J. Binks[1,2*]**

[1] Photon Science Institute, University of Manchester, M13 9PL, UK
[2] School of Physics and Astronomy, University of Manchester, M13 9PL, UK
[3] School of Chemical Engineering & Analytical Science, Univ. of Manchester, M13 9PL, UK
[4] National Graphene Institute, University of Manchester, M13 9PL, UK
[5] Manchester Institute of Biotechnology, University of Manchester, M13 9PL, UK
[6] Department of Materials Science and Engineering, National University of Singapore, Singapore, 117575, Singapore
*To whom correspondence should be addressed

**ABSTRACT** The pH dependence of emission from graphene oxide is believed to be due to the protonation of surface functional groups. In this study we use transient absorption spectroscopy to study the sub-picosecond charge dynamics in graphene oxide over a range of pH values, observing dynamics consistent with an excited state protonation step for pH < 9.3. The timescale of this process is ~ 1.5 ps, and a corresponding change in recombination dynamics follows. A broad photo-induced absorption peak centred at 530 nm associated with excited state protonation is also observed.

Email: david.binks@manchester.ac.uk

## 1. INTRODUCTION

Graphene oxide (GO), which is often used as a precursor for the chemical synthesis of graphene[1], has gained interest itself due to its photoluminescence (PL) properties and biocompatibility. The mechanisms behind GO PL are particularly interesting as they differ from those found in 2D semiconductors with visible optical band-gap emission such as transition metal dichalcogenides[2,3]. Functionalization of graphene to form graphene oxide creates a band-gap, but this gap is small (~0.6 eV[4]), yet GO emission is observed ranging from the blue to near infra-red[5,6].

The emission from GO is widely agreed to be due to molecular-type PL associated with confined 'islands' of graphitic $sp^2$ regions surrounded by functionalized $sp^3$ regions on the GO surface[4–13]. The size of the $sp^2$ regions depends on the nature and amount of functionalization, which strongly affects emission energy as the molecule-like orbital energy levels around these islands changes[14]. Hence, the PL has been found to depend on various factors, including the functional group used[15,16] and the degree of reduction[16]. In particular, contributing to the PL are transitions associated with the -OH and -COOH groups at short and long wavelengths respectively[12]. The -COOH groups are mostly deprotonated to -COO$^-$ but in acidic conditions undergo reprotonation upon excitation as illustrated in Figure 1. This excited state protonation enhances long wavelength emission, resulting in pH dependent emission bands which are commonly reported in the literature[10–12]. The pH dependent behavior of GO can be utilized in a number of ways, but has particular use in biosensing and drug release, due to its biocompatibility and strong affinity to biological molecules such as single-strand DNA[17,18]. This interaction is due to π-π interactions with ring structures in the nucleic acid bases[19], but the conformal change associated with surface group protonation modifies the interaction and has led to the use of GO as a targeted drug delivery system[20-23]. In order to fully control this release process, a good understanding of the protonation step must be developed. This process occurs on a picosecond timescale and so ultrafast experiments are required to study it.

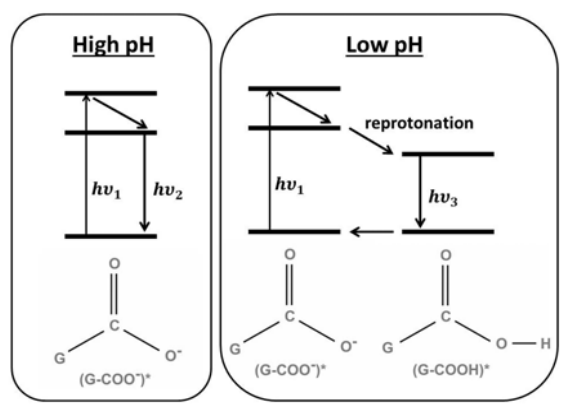

Figure 1. Schematic illustration of the photoluminescence process. At high pH, G-COO$^-$ is excited to (G-COO$^-$)* by a photon of energy $h\nu_1$ and then relaxes to a lower lying level before emitting a photon of energy $h\nu_2$. However, at low pH, excited state reprotonation occurs forming (G-COOH)*, which emits a photon of energy $h\nu_3$.

Previous studies of charge dynamics in GO have presented the results of PL lifetime measurements, transient absorption and femtosecond upconversion spectroscopy. These studies all show transient PL and absorption decays with a multi-exponential form, consistent with multiple populations of emitting species, which can be characterized by 3 or 4 time constants with values ranging from 0.17 ps to 1680 ps[4,24,25]. Some pH dependent work has been reported: Zhang et al. report two bands of PL centred at 440 nm and 650 nm, with the longer wavelength band only present at lower pH levels where protonation can more readily take place[11]; Du *et al* have detailed the concentration dependence of the pH-driven spectral shift in emission[26]; and Konkena et al temporally resolved these spectral shifts on a nanosecond timescale[27]. However, transient studies of the pH-dependence of GO emission have yet to be carried out with sufficient resolution to establish the timescale of the reprotonation process and thus its potential impact on the time response of its applications.

In this study we use transient absorption spectroscopy to observe and study the pH-dependent time-evolution of the protonation step in GO. This is complemented by steady state absorption and PL spectroscopy, and transient PL measurements.

## 2. METHODS.

A GO suspension was prepared by a 4-hours sonication-assisted exfoliation of graphite oxide powder made by a modified Hummers method. The resulting suspension was centrifuged 3 times at 8,000 rpm to remove the multilayer GO flakes. To obtain atomic force microscope (AFM) images, the graphene oxide suspension (0.1mg/mL) was diluted 1000 times and deposited onto a silicon/silicon oxide wafer (2 x 2 cm) by drop casting. The sample was dried overnight in the oven at 50 °C. The AFM measurements were carried out using Veeco Dimension 3100 under tapping mode. The data was analyzed by NanoScope Analysis software. 0.1 mg/mL GO samples with different pH value were prepared by adding KOH or HCl solutions to the GO suspension according to previous reports[10,28]. The pH values of the suspensions were determined using a pH titration monitored by a Mettler Toledo pH meter (SevenExcellence S470 Benchtop, pH accuracy~0.001). Magnetic stirring was constantly applied to ensure the homogeneity of the dispersion. We note that there is no visible separation of graphitic material or oxidative debris as referenced by Coleman et al.[29] in our samples; all samples studied remain stable in suspension over many weeks.

Steady-state absorption and photoluminescence spectroscopy were performed using a Perkin Elmer Lambda 1050 UV/Vis/IR spectrophotometer and a Horiba Fluorolog-3 spectrofluorometer respectively. Excitation at 400 nm was used in the photoluminescence experiments.

Photoluminescence decay transients were obtained using time-correlated single photon counting (TCSPC). A mode-locked Ti:Sapphire (Spectra-Physics Mai Tai) generated 100 fs, 800 nm pulses at 80 MHz. The pulse rate was slowed to 4 MHz using a pulse picker (APE Pulse Select) and frequency doubled to 400 nm. TCSPC counting electronics (Edinburgh Instruments T900) were used to collect lifetime measurements for photoluminescence at 580 nm. Emission was collected via a monochromator (Horiba, SPEX 1870C) and microchannel plate (Hamamatsu R3809U-50). To ensure single photon counting, the excitation/emission count rate ratio was kept below 1%.

A broadband ultrafast pump-probe transient absorbance spectrometer (Helios, Ultrafast Systems LLC) was used to collect transient absorption data. The pump and probe beam were created by a Ti:sapphire amplifier system (Spectra Physics Solstice Ace) producing 800 nm pulses at 1 kHz with a 100 fs pulse duration. This beam was then split to form pump and probe beams. The pump portion was passed through an optical parametric amplifier (Topas Prime) to generate 300 nm excitation pulses with a FWHM (full width at half-maximum) of ~10 nm. The excitation beam at the sample had a beam diameter of ~200 µm and average power of 470 µW. The probe portion was passed through a rastered $CaF_2$ crystal to form a white light continuum, and absorption changes were monitored between 330–700 nm. Data was collected from -3 ps to 2 ns, with a time resolution of ± 0.22 ps, measured by the Raman response in water. Data at 600 nm is dominated by the second order diffraction of scattered pump light within the spectrometer, and so has been removed.

## 3. RESULTS AND DISCUSSION

AFM measurements of the GO suspension found that the lateral size (square root of the area of the flake) of the flakes varied from 0.4 to 1.2 µm. The distribution of sizes had an average of 0.75 µm and a standard deviation of 0.10 µm; an AFM image and size histogram are given in the Supporting Information (Figure S1).

The absorbance spectra for GO colloidal suspensions under acidic and alkaline conditions are compared in Figure 2a, and shown for a range of pH values in the Supporting Information (Figure S2a). In each spectra, there is a peak at 230 nm and shoulder at ~300 nm associated with π-π* transitions of C-C and C=C bonds in $sp^2$ regions and n-π* of C=O bonds in $sp^3$ regions, respectively. While we see no significant spectral shift in these features with pH, the attenuation coefficient at wavelengths longer than 230 nm is significantly higher for basic suspensions as has been previously reported[10].

In contrast to the small changes in the absorption spectrum, we observe large changes in the PL spectrum as the pH is increased, as shown in Figure 2b and in the Supporting Information (Figure S2b) for excitation at 400 nm; an example spectrum for excitation at 300 nm is also shown in Supporting Information (Figure S3). A clear difference is seen between low (< 9.3) and high (> 9.3) pH, with broad emission bands evident centred at ~700 nm and ~520 nm, respectively. This transition between emission bands due to pH has been reported previously and is attributed to radiative recombination occurring through the –OH or -COOH surface groups[10–12,30]. At high pH, a blue band arises from –OH transitions, whereas in more acidic media PL is dominated by excited state transitions in the protonated –COOH group, leading to a broad red emission pathway. We note that in our data the blue band may still be present below pH 9.3 but is hidden by the strong red emission, which can be seen more clearly in Figure S4. The excitation spectra for the pH 4.2 and 9.3 samples are given in the Supporting Information (Figure S5) and broadly match the absorbance spectra in Figure 2a; there is no indication of the effect of any impurities on the photoluminescence.

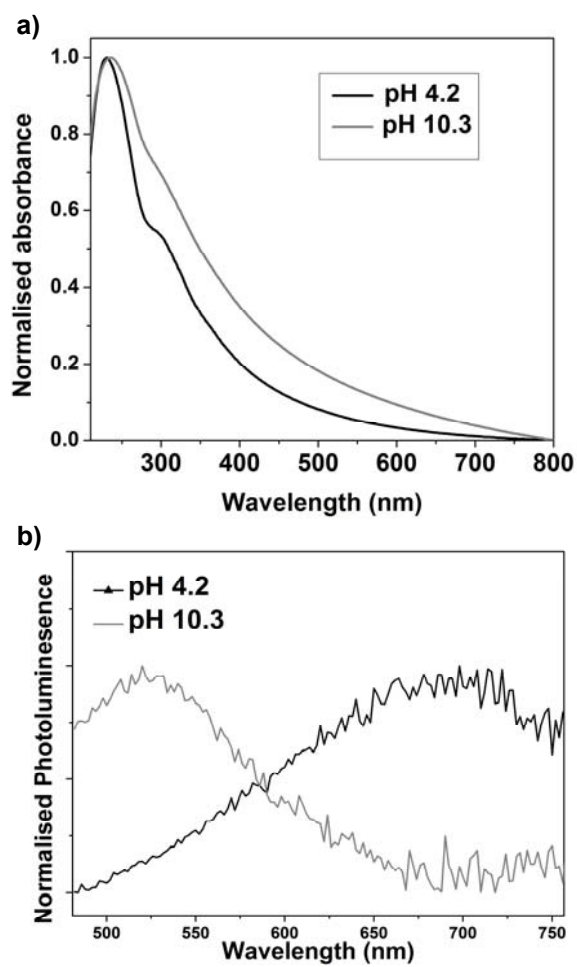

Figure 2. pH dependence of (a) absorbance spectra, normalised to the π-π* absorbance feature at 230 nm, (b) normalised photoluminescence spectra for excitation at 400 nm.

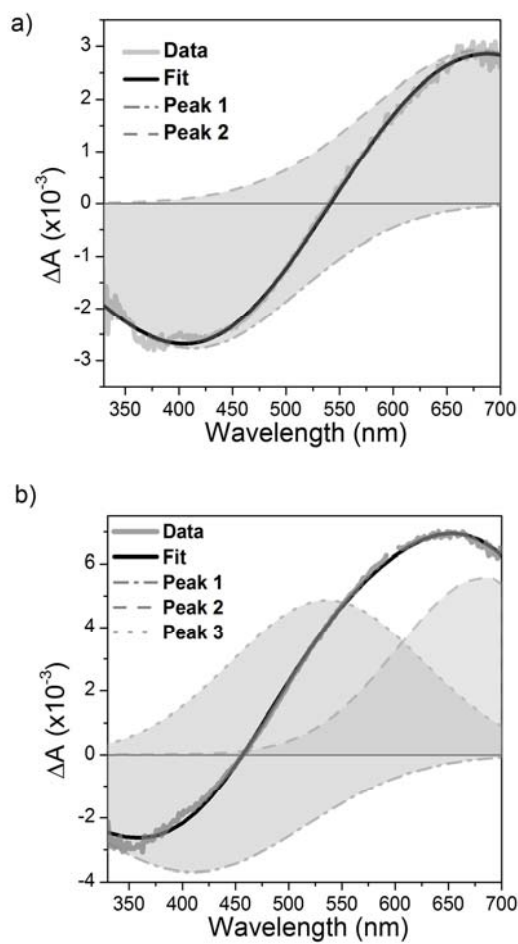

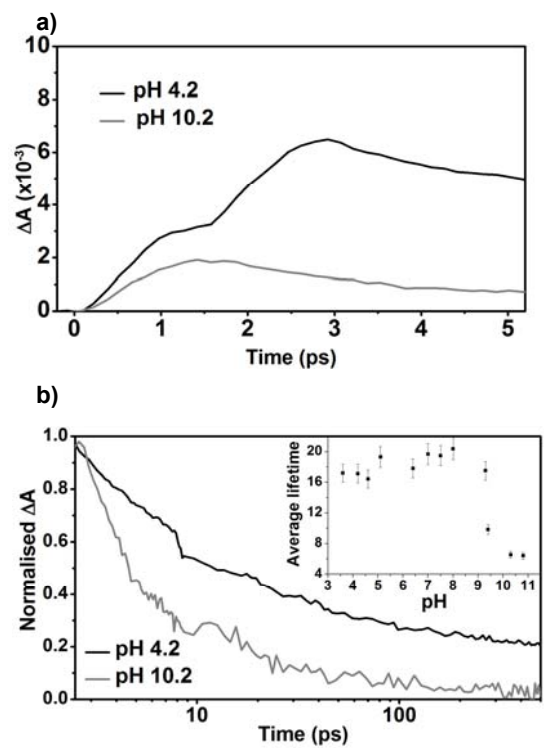

Figure 4 Differential absorption transients showing (a) stepwise growth under acidic conditions which disappears when the dispersion becomes sufficiently alkali (i.e. pH>9.3), and (b) more rapid decay under alkali conditions. Inset: the average decay time as a function of pH.

Figure 3 Differential absorption, $\Delta A$, spectra produced by pumping at 300 nm for a pH of a) 10.2, and b) 4.2. Each spectrum is taken at its respective maximum amplitude, around 2 ps after excitation. Also shown are fits to the spectra composed of a) 2 or b) 3 Gaussian peaks.

To explore this phenomenon in more detail, we studied the PL lifetimes at 580 nm where both PL features have significant intensity. The resultant decay transients are shown in the Supporting Information (Figure S2b inset). At all pH values, the majority of the PL decay is very rapid and likely instrument response limited. However, for highly alkaline samples (pH>9.3) a slower component with a lifetime of 1.8±0.3 ns is seen, which is only weakly present for the other pH values. An increased PL lifetime for highly alkaline GO has been reported by Zhang et al[11], although we note that our observed emission lifetimes are shorter lived and the transition in behavior occurs at a higher pH.

These PL decay transients show that the charge dynamics occur largely on a sub-nanosecond timescale and so ultrafast transient absorption spectroscopy was used to study them further. Differential absorption, $\Delta A$, spectra under acidic and alkaline conditions are compared in Figure 3, with spectra for a range of pH values given in the Supporting Information (Figure S6). Each spectrum is composed of a broad bleach feature at short wavelengths and photoinduced absorption feature at longer wavelengths. This latter feature is similar to that reported by Shang et al. and was attributed to excited state absorption, although they did not probe below 530 nm to the spectral region where we observe the bleach feature[4]. For pH values >9.3, the differential spectra are well described by a fit comprising two Gaussian peaks: one (peak 1) corresponding to a bleach centred at about 400nm and another (peak 2) corresponding to a photo-induced absorption centred at about 650 nm.

However, for pH values <9.3, the differential spectrum is described by a fit comprised of peaks 1 and 2 plus an additional photo-induced absorption centred at about 530 nm (peak 3). This pH-dependent photo-induced feature is attributed to the formation of G-COOH from G-COO$^-$ via excited state protonation.

To further understand this process, its time evolution was studied. Figure 4 shows the rise and decay of $\Delta A$ at 610 nm, which will have contributions from both peak 2 and 3. There is again a stark change at pH 9.3. For pH > 9.3, the maximum $\Delta A$ is achieved at ~1.5 ps following monotonic growth. In contrast, for pH < 9.3 a stepwise growth is observed with the first step occurring over ~1.5 ps and the second also taking ~1.5 ps. This behavior is consistent with the production of an additional species whose formation is dependent on the creation of a precursor. The formation of (G-COOH)* by excited state protonation requires the initial creation of (G-COO$^-$)*. The data suggests that the creation of (G-COO$^-$)* occurs over the first ~1.5 ps, and is followed by a protonation step also taking about ~1.5 ps. This second process is suppressed in highly alkaline conditions and therefore this two-step process is not seen above pH 9.3.

The subsequent $\Delta A$ decay transients are shown in Figure 4b for typical acidic and alkaline dispersions, with the decays for a range of pH values given in the Supporting Information (Figure S7). The decays observed for pH>9.3 are significantly more rapid than those observed for more acidic dispersions. A similar increase in the decay rate for higher pH values has also been reported for carbon nanodots[31]. Each of the decays can be described by a triexponential function with lifetimes $\tau_1$=2.0±0.1 ps, $\tau_2$=15±1 ps and $\tau_3$=140±10 ps, which are values similar to previous reports[4,24,25]. However, the amplitude of component associated with $\tau_1$ is greater for pH>9.3; this can be seen from the inset to Figure 4b which shows the average lifetime, $\bar{\tau}$, for a range of pH values. The value of $\bar{\tau}$ is calculated from the individual lifetimes weighted by the amplitude of the corresponding component; these amplitudes are given in the Supporting Information (Figure S8). The change in decay rate at pH=9.3 is consistent with the formation of a new species of functional group for pH values less than this i.e. the excited state protonation of (G-COO$^-$)* to (G-COOH)*.

In this study we see an abrupt change in behavior at pH=9.3, whereas other reports have witnessed or predicted the change at pH levels ranging from 5 to 9.5 [10,11]. This difference in the critical pH level is attriubuted to the sensitivity of GO's optical properties to surface functionalization which in turn depends on fabrication techniques[7]. To fully utilize this pH dependent behavior, both understanding of and control over the onset of excited state protonation must be gained.

## 4. CONCLUSION.

We find that the optical properties and charge dynamics of dispersions of GO undergo an abrupt change in behavior at pH 9.3. These results are consistent with the excited state protonation of (G-COO$^-$), which leads to a pH dependent shift in the absorbance and PL spectra, and in the PL decay transients. This excited-state protonation also produces a broad photo-induced absorption peak centred at 530 nm. This peak exhibits stepwise growth that allows the time-scale of the protonation process to be determined as about 1.5 ps. Additionally, the recombination dynamics are affected by the onset of excited state absorption, with a increase in fast recombination pathways above pH 9.3. This understanding of the rate and effects of the protonation process will aid the exploitation of the pH-dependence of GO for applications such as biosensing and drug-release. In particular, the picosecond timescale of protonation revealed by this study means that it is very unlikely to limit the response time of applications based on GO pH-dependence.

**ASSOCIATED CONTENT**
**Supporting information**
The Supporting Information is available free of charge on the ACS Publications website.

Atomic force microscopy image and size histogram; absorbance spectra and PL spectra for excitation at 400 nm; example PL spectrum for excitation at 300 nm; PL decay transients; excitation spectra for pH 4.2 and 9.3 samples; differential absorption spectra and transients; PL decay component amplitudes.

**AUTHOR INFORMATION**
**Corresponding Author**
*E-mail: david.binks@manchester.ac.uk

**Notes**
The authors declare no competing financial interest.

**ACKNOWLEDGEMENTS**
This work was funded by the EPSRC under grants EP/L01548X/1 and EP/N010345/1. RRN and KSN are grateful to Graphene Flagship, Royal Society and European Research Council (contracts 679689 and 319277 Hetero2D). Transient absorption measurements were performed at the Ultrafast Biophysics Facility, Manchester Institute of Biotechnology, as funded by BBSRC Alert14 Award BB/M011658/1. The data associated with this paper are openly available from Mendeley data: doi:10.17632/n9tvcwccf5.1

# The pH Dependence of Ultrafast Charge Dynamics in Graphene Oxide Dispersions

**Georgia Kime[1,2], Kai-Ge Zhou[3,4], Samantha J. O. Hardman[5], Rahul R. Nair[3,4], Konstantin Sergeevich Novoselov[2], Daria V. Andreeva[6], and David J. Binks[1,2*]**

[1] Photon Science Institute, University of Manchester, M13 9PL, UK
[2] School of Physics and Astronomy, University of Manchester, M13 9PL, UK
[3] School of Chemical Engineering & Analytical Science, Univ. of Manchester, M13 9PL, UK
[4] National Graphene Institute, University of Manchester, M13 9PL, UK
[5] Manchester Institute of Biotechnology, University of Manchester, M13 9PL, UK
[6] Department of Materials Science and Engineering, National University of Singapore, Singapore, 117575, Singapore
*To whom correspondence should be addressed

**S1. Atomic force microscope measurements.**

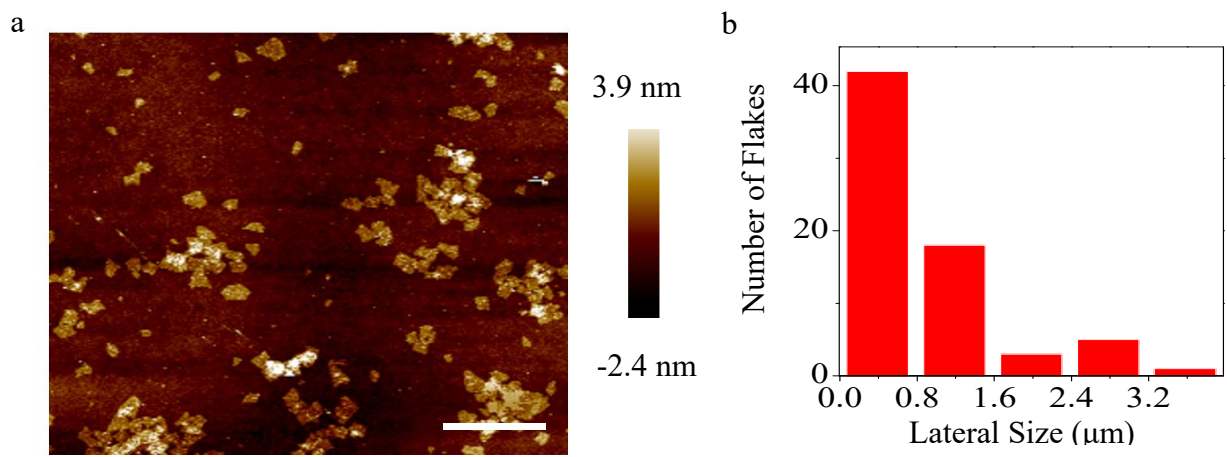

Figure S1 a) The AFM height image of as-prepared graphene oxide flakes deposited on SiO2/Si substrate. Scale bar: 4 µm; b) the lateral size distribution of the graphene oxide flakes. The lateral sizes of flakes were estimated by taking the square root of the area of each flake measured with the NanoScope Analysis software.

**S2. Absorbance and photoluminescence spectra & photoluminescence decay transients.**

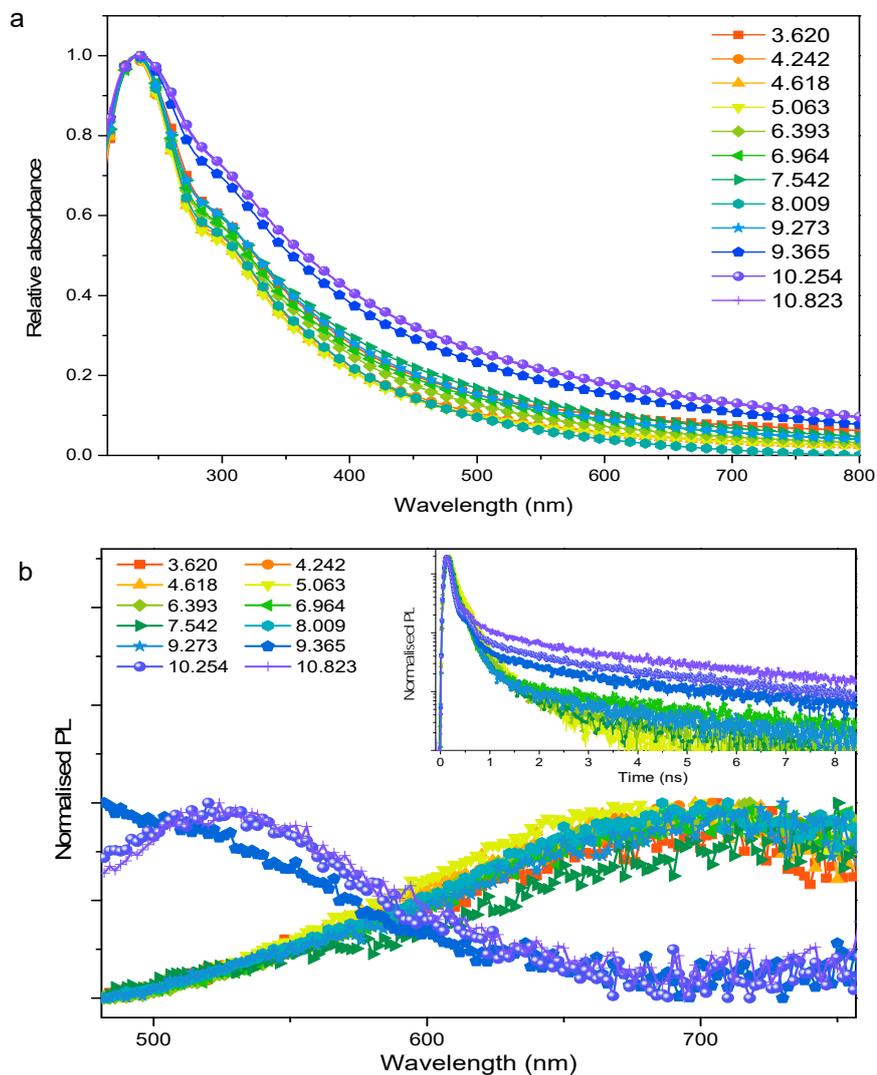

Figure S2 pH dependence of (a) absorbance spectra, normalised to the π-π* absorbance feature at 230 nm, (b) normalised photoluminescence spectra for excitation at 400 nm, and (inset) photoluminescence transient measurements for emisson at 580 nm.

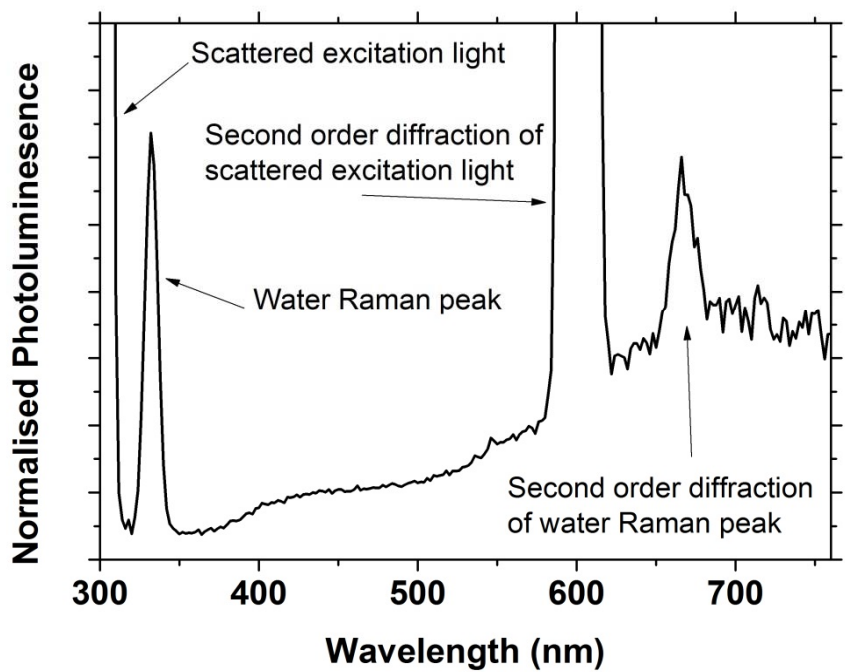

Figure S3. Spectrum for the pH=4.2 GO dispersions excitated at 300 nm showing broad photoluminescence, and Rayleigh and Raman scattered excitation light that has undergone both first and second order diffraction within the spectrometer.

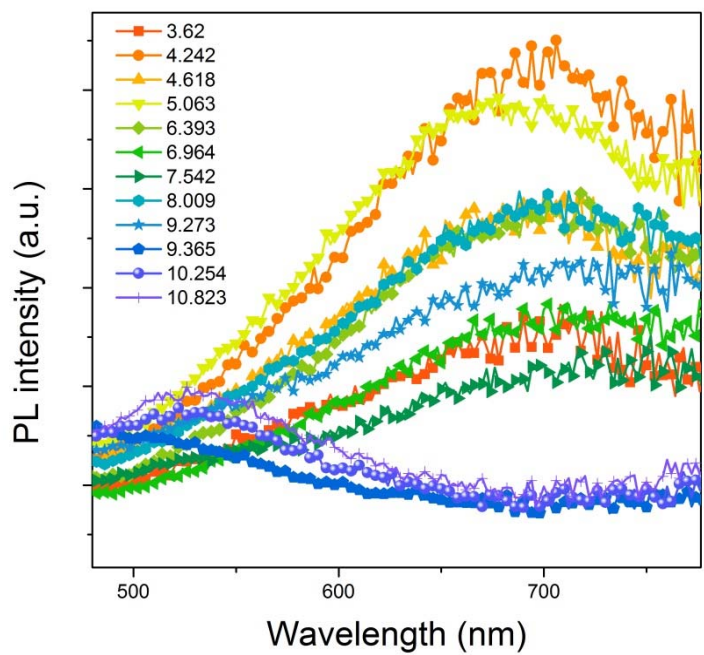

Figure S4. Steady-state photoluminescence of dispersions of GO at a range of pH levels, for excitation at 400 nm and without normalization.

## S3. Excitation spectra.

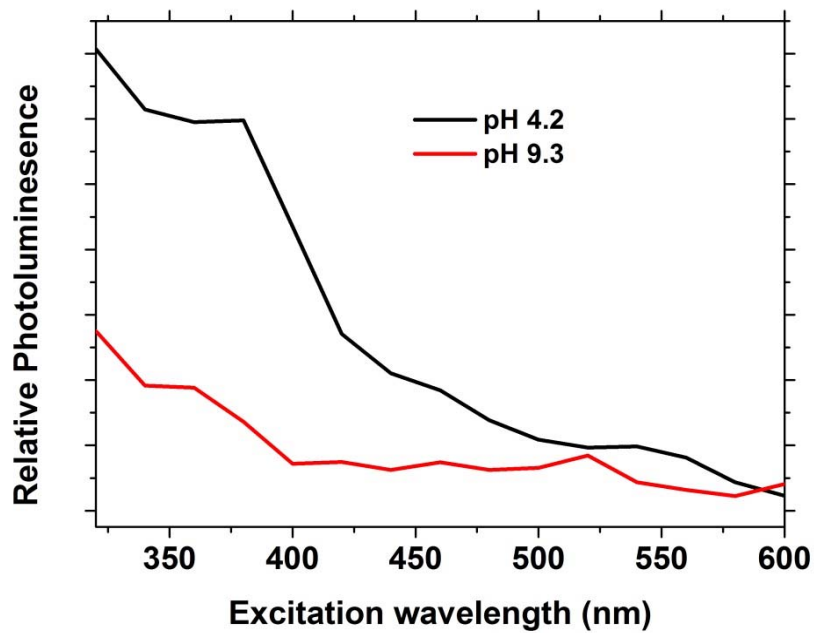

Figure S5. Excitation spectra for the pH 4.2 and pH 9.3 samples for a detection wavelength of 620 nm.

## S4. Differential absorption spectra.

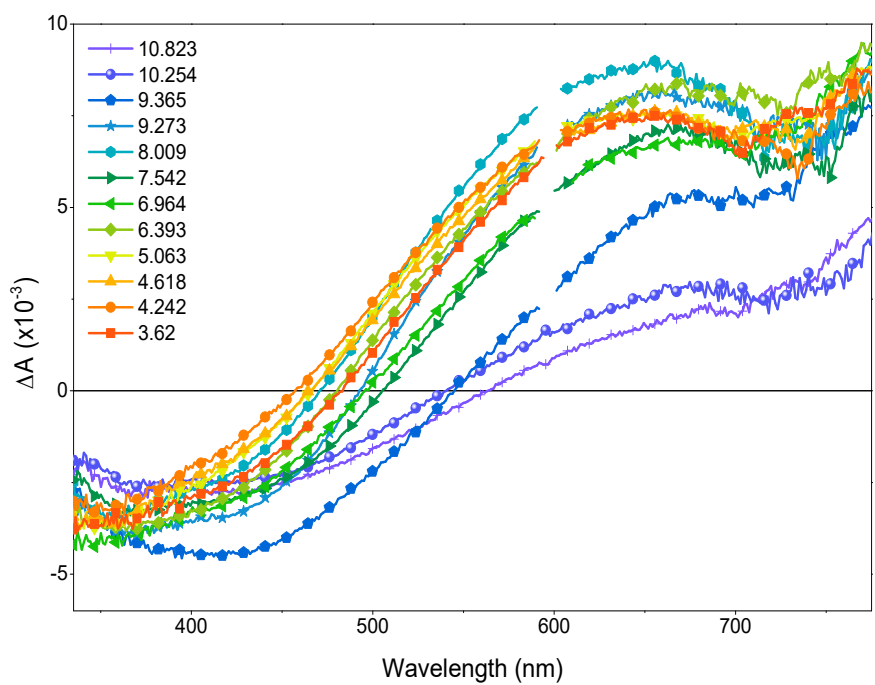

Figure S6 Differential absorption, ΔA, spectra between 330 – 750 nm following excitation at 300 nm. Each spectrum is taken at its respective maximum amplitude, around 2 ps after excitation.

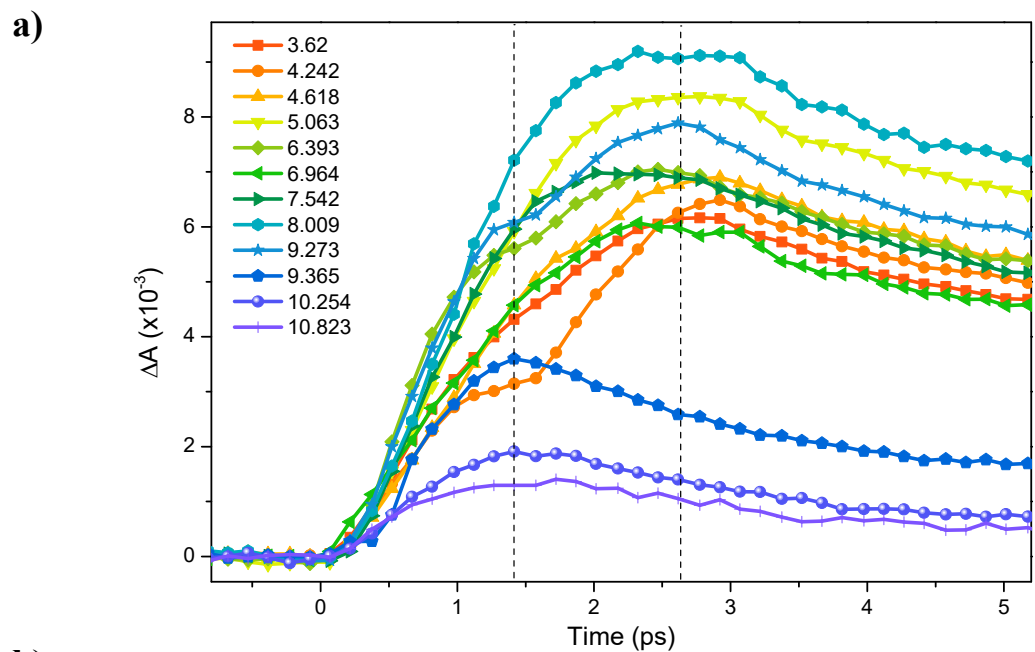

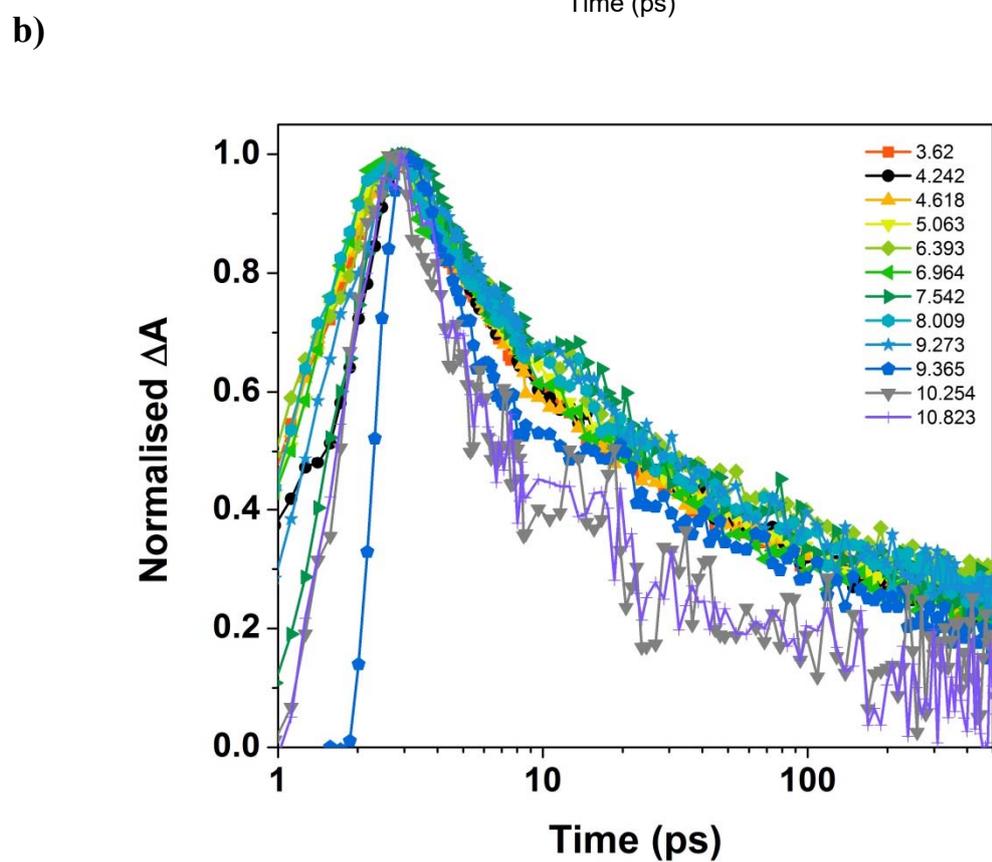

Figure S7 Differential absorption transients showing (a) stepwise growth fro pH<9.3 and monotonic growth for pH>9.3, and (b) more rapid decay for pH>9.3.

## S5. Decay component amplitudes.

Globally fitting tri-exponential decays to figure 4b yields time constants $\tau_1$ = 2.0±0.1 ps, $\tau_2$ = 15±1 ps and $\tau_3$ = 140±10 ps. The variation in amplitude of each component with pH is plotted in Figure S, in which it can be seen that the amplitude associated with $\tau_1$ increases significantly for pH>9.3, with the amplitudes associated with the other time constants drops at the same point.

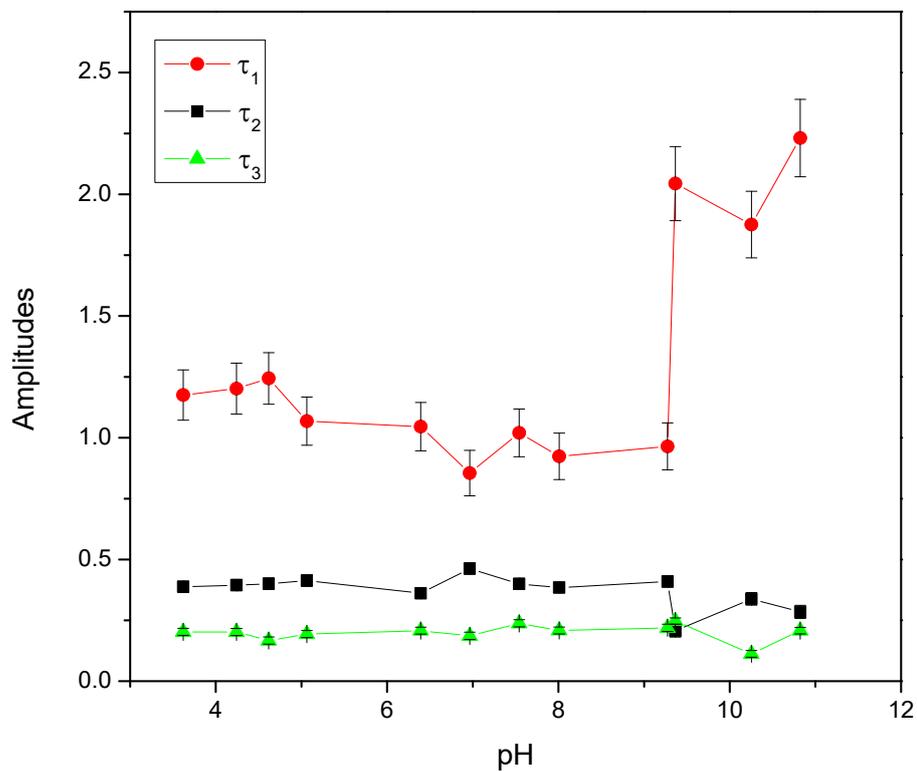

Figure S8. Amplitudes associated with the three exponential time constants ($\tau_1$, $\tau_2$, $\tau_3$) fitted to the transient absorption decay curves of GO dispersions at a range of pH levels.